\documentclass[11pt]{article}
\usepackage{amsmath}

\title{Positiveness and Pauli exception principle \\
in raw Bloch equations for quantum boxes}
\author{Brigitte Bid\'egaray-Fesquet\\
\small Universit\'e de Grenoble and CNRS UMR 5224\\
\small Laboratoire Jean Kuntzmann \\
\small BP 53, 38041 Grenoble Cedex 9, France}

\def\bfr{{\bf r}}

\def\bfE{{\bf E}}
\def\bfM{{\bf M}}
\def\bfN{{\bf N}}
\def\Ic{{I^{\rm c}}}
\def\Iv{{I^{\rm v}}}

\def\Id{{\rm Id}}
\def\Tr{\operatorname{Tr}}
\def\Re{\operatorname{Re}}
\def\diag{\operatorname{diag}}

\def\ci{{\rm i}}

\begin{document}

\maketitle

\begin{abstract}
The aim of this paper is to derive a raw Bloch model for the interaction of light with quantum boxes in the framework of a two-electron-species (conduction and valence) description.
This requires a good understanding of the one-species case and of the treatment of level degeneracy.
In contrast with some existing literature we obtain a Liouville equation which induces the positiveness and the boundedness of solutions, that are necessary for future mathematical studies involving higher order phenomena.
\end{abstract}

\section{Introduction}

Bloch equations are a very common model to describe the time evolution of a system of electrons in different contexts such as gas of electrons, glasses, crystals.
Mathematical results have been obtained in these contexts which ensure in particular that the variable of Bloch equations, namely the density matrix, keeps through the time evolution some properties, such as hermicity and positiveness (of diagonal elements and of the matrix as an operator) \cite{Bidegaray01b,Bidegaray-Bourgeade-Reignier01}.
In the above contexts, the indices in the density matrix are integers which distinguish between the different electron levels.
Diagonal entries model the population of the levels, while off-diagonal entries
model coherences between two levels.

Bloch equations have also been derived for quantum wells \cite{Hess-Kuhn96a,Kuhn-Rossi92}. 
They look very similar but the variables are different. 
In the quantum well context, we distinguish between two species of electrons
(valence and conduction electrons) and variables are indexed by (two-dimensional) wave vectors. 
Electrons can interact directly only if they have the same wave vector, which induces intra-band coherences to be zero and valence and conduction electrons to be coupled two-by-two. 
This is not a mathematical problem since at first order, i.e. without considering relaxation effects, such a system is a juxtaposition of two-level models.

The model for quantum wells has been extended to quantum boxes \cite{Gehrig-Hess02}.
Due to the three-dimensional confinement of electrons, variables are now again indexed by integers. 
Valence and conduction electron levels are not coupled two-by-two any more and any valence electron can interact with any conduction electron.
But in the propounded model intra-band coherences are still considered as zero.
This destroys the Liouville structure of the system which in other respects resembles much the gas model.
Another puzzling point is the fact that levels are supposed to have any degree of degeneracy. 
This is not much coherent with the Pauli exception principle that electrons (which are fermions) are supposed to follow.
This induces also mathematical and modelling problems which we also address here.

In this paper we are interested in the derivation of a raw Bloch model for a quantum box system.
This model will be derived carefully from computations involving creation and annihilation operators for both species of electrons. 
Our goal is to obtain a model with appropriate mathematical properties in order to be able to prove the conservation of certain physical properties through time. 
Our study will be restricted to the leading order terms in the Hamiltonian, leaving extra contributions such as Coulomb effect or electron--phonon interaction to future works. More precisely this is not a restricted study but the first step towards a broader study, since the properties proved on the first order terms are necessary to study the higher terms.

The outline of this paper is as follows. 
In Section \ref{Sec_1fermion} we derive carefully a one-species Bloch model such as the one for gazes. 
The structure of this model is analysed. 
Positiveness results are recalled and boundedness of the populations is discussed.
This motivates Section \ref{Sec_degenerate} where the problem of degeneracy is thoroughly addressed.
Section \ref{Sec_2fermion} is devoted to our central problem, i.e. the derivation of a two-species Bloch model.
This model has the desired mathematical properties.
The details of the computations and an analogy with a boson model are postponed to various appendices.

\section{The case of one species of electrons}
\label{Sec_1fermion}

\subsection{Commutators and Heisenberg equation}

Let $A$ and $B$ be two operators, we define their commutator by
\begin{equation*}
[A,B]=AB-BA  
\end{equation*}
and their skew-commutator by
\begin{equation*}
\{A,B\}=AB+BA.
\end{equation*}
For an operator $A$, we will define the associated observable $\langle A\rangle$
\begin{equation*}
\langle A \rangle = \Tr(S_0 A)
\end{equation*}
by averaging with respect to the initial state density $S_0$ of the system.
If the system is described by an Hamiltonian $H$, the time-evolution for this observable is given by the Heisenberg equation
\begin{equation*}
\ci\hbar \partial_t \langle A \rangle = \langle [A,H] \rangle. 
\end{equation*}

Bloch equations are the Heisenberg equation of motion when the observable is the density matrix.

\subsection{Operators and commutation rules}

We suppose there is only one species of electrons, like in the case of gas atoms, and that levels are indexed by integers $i\in I$. 
For the $i$th level, we define the creation and annihilation operators $c_i^\dag$ and $c_i$. 
Electrons are fermions and should respect the Pauli exception principle.
The corresponding skew-commutation rules are
\begin{equation*}
\{c_i,c_j\}=\{c_i^\dag,c_j^\dag\}=0, \hspace{1cm} \{c_i,c_j^\dag\} = \delta_{i,j}.
\end{equation*}
This implies in particular that
\begin{equation*}
c_i c_i^\dag c_i = c_i, \hspace{1cm} c_i^\dag c_i c_i^\dag = c_i^\dag, \hspace{1cm} c_i c_i = c_i^\dag c_i^\dag = 0,
\end{equation*}
this last equality meaning clearly that it is impossible to create twice or annihilate twice the same electron. 
This is the Pauli exception rule.
Let $A$ and $B$ be the products of $n_A$ and $n_B$ such operators which all skew-commute two-by-two, then 
\begin{equation*}
[A,B] = 
\begin{cases}2AB & \textrm{if $n_A$ or $n_B$ is odd}, \\
0 & \textrm{else.}
\end{cases}
\end{equation*}
Since we only use two-operator observables, we will mostly always be in the second case.

\subsection{Hamiltonian and observables}

We define the Hamiltonian as the sum of a part due to the electrons and only, and a part due to their interaction with an external electric field $\bfE(t)$: 
$H=H^{\rm e} + H^{\rm L}$ with
\begin{equation*}
H^{\rm e} = \sum_{k\in I} \epsilon_k c_k^\dag c_k
\end{equation*}
where $\epsilon_k$ is the energy of an electron in the $k$th level and 
\begin{equation*}
H^{\rm L} 
= \frac12 \sum_{(k,l)\in I^2} ( \bfE(t)\cdot\bfM_{kl} c_k^\dag c_l + \bfE^*(t)\cdot\bfM_{kl}^* c_l^\dag c_k),
\end{equation*}
where $\bfM_{kl}$ is an element of the dipolar moment matrix $\bfM$.
The $1/2$ coefficient is due to the fact that interactions are counted twice in this sum. 
The dipolar moment $\bfM_{kl}$ can be expressed as $\langle \psi_l^*|\bfr|\psi_k \rangle$ in terms of the wave functions associated to each level and the local position vector $\bfr$.
This implies in particular that the case $k=l$ does not contribute to the sum since $\bfM_{kk}=0$. 
We will also use the fact that $\bfM$ is hermitic~: $\bfM_{kl}^*=\bfM_{lk}$.

The observable we are interested in is the density matrix, which elements are the $\rho_{ij}=\langle c_j^\dag c_i\rangle$. It is clear that this matrix is Hermitian.
In the computations, we distinguish between diagonal terms of this matrix also called populations $\rho_{ii}=\langle c_i^\dag c_i\rangle$ and off-diagonal terms, $\rho_{ij}$, $i\neq j$ also called coherences.

\subsection{Computation of the raw Bloch equations}

We have to compute the commutators of the density matrix with both Hamiltonians.
The details of the computations are given in Appendix \ref{App_1fermion}.
We obtain
\begin{equation*}
\langle[c_j^\dag c_i,H^{\rm e}]\rangle = (\epsilon_i-\epsilon_j) \rho_{ij},
\end{equation*}
and
\begin{equation*}
2\langle[c_j^\dag c_i,H^{\rm L}]\rangle
= \Re\bfE(t) \cdot \sum_{k\in I} (\bfM_{ik} \rho_{kj} - \bfM_{kj} \rho_{ik}).
\end{equation*}
Gathering these results and the Heisenberg equation, raw Bloch equations read
\begin{equation*}
\ci\hbar\partial_t \rho_{ij} = (\epsilon_i-\epsilon_j) \rho_{ij} 
+ \Re\bfE(t) \cdot \sum_{k\in I} (\bfM_{ik} \rho_{kj} - \bfM_{kj} \rho_{ik}).
\end{equation*}
If we denote $E=\diag(\{\epsilon_i\}_{i\in I})$, this can be cast as a Liouville equation
\begin{equation}
\label{Eq_Bloch1}
\ci\hbar\partial_t \rho = [V(t),\rho], 
\hspace{1cm} \textrm{ where }V(t) = E + \Re\bfE(t)\cdot\bfM.
\end{equation}
In Appendix \ref{App_boson} we show that considering electrons as bosons leads to exactly the same equation, although the intermediate computations are different.
This is not true any more if you consider extra contributions to the Hamiltonian such as Coulomb effect or electron--phonon interaction, but this is an other story.

\subsection{Positiveness and trace results}

Equation \eqref{Eq_Bloch1} clearly preserves the Hermitian structure of $\rho$.
We can give an exact solution to equation \eqref{Eq_Bloch1}, namely
\begin{equation*}
\rho(t) = \exp\left(-\frac{\ci}{\hbar} \int_0^t V(\tau) d\tau\right) \rho(0) 
\exp\left(\frac{\ci}{\hbar} \int_0^t V(\tau) d\tau\right).
\end{equation*}
This evolution equation preserves positiveness.
Indeed, if $\rho(0)$ is a non negative operator, that is for all $X\in l^2(I)$, 
$X^*\rho(0)X\geq0$, then
\begin{eqnarray*}
X^*\rho(t)X 
& = & X^*\exp\left(-\frac{\ci}{\hbar} \int_0^t V(\tau) d\tau\right) \rho(0) 
\exp\left(\frac{\ci}{\hbar} \int_0^t V(\tau) d\tau\right) X \\
& = & Y^*\rho(0) Y, \hspace{1cm} \textrm{ where } Y=\exp\left(\frac{\ci}{\hbar} \int_0^t V(\tau) d\tau\right)X.
\end{eqnarray*}
Therefore $X^*\rho(t)X \geq 0$.
A corollary of this is that for all time $t$, and all level $j$, $\rho_{jj}(t)\geq0$.

The total population is the sum of populations, which can be described by the trace of the density matrix.
Since commutation is a trace preserving operation, we have
\begin{equation*}
\ci\hbar\partial_t \Tr\rho = \Tr[V(t),\rho] = 0.
\end{equation*}
Hence $\Tr \rho(t)=\Tr \rho(0) \equiv N_{\rm tot}$.
The raw Bloch equations are trace preserving. 
In other words the total number of electrons is preserved through the time evolution. 
This together with non negativeness ensures that for all time $t$
\begin{equation*}
0 \leq \rho_{jj}(t) \leq N_{\rm tot}.
\end{equation*}

We want also to know whether each level remains bounded by 1, which would be the expression of the Pauli exception principle.
This is indeed the case.
If we set $\tilde \rho=\Id-\rho$, where $\Id$ is the identity matrix with the same dimensions as $\rho$, then
\begin{eqnarray*}
[V,\Id-\rho] & = & [V,\Id] - [V,\rho] = - [V,\rho], \\
\partial_t (\Id-\rho) & = & - \partial_t \rho.
\end{eqnarray*}
Therefore $\tilde \rho$ is also solution to the raw Bloch equations, and in particular $\tilde \rho_{ii}(t) \geq 0$, which means $\rho_{ii}(t) \leq 1$.

\section{Degenerate levels}
\label{Sec_degenerate}

In \cite{Gehrig-Hess02}, levels are supposed to be degenerate, with a degeneracy order that depends on the levels. 
In a fermion description levels cannot be degenerate since each level population is bounded by 1.
This would \textit{a priori} be possible in a boson description.
Let us first investigate this situation. 

The same proof as above shows that if $\rho$ is solution to the raw Bloch equations and the level populations are initially bounded by the same constant (say $d$), then $\rho_{ii}(t) \leq d$.
To derive a $d$-degenerate raw Bloch model, it is therefore possible to consider electrons as bosons and impose a condition on the initial data. 
But this solution is not very satisfactory from two points of view. 
First this cannot be extended to the case when the degeneracy depends on the level, i.e. the $i$th level is $d_i$-degenerate. 
Second if we want to look at other contributions in the Hamiltonian, bosonic and fermionic commutation rules do not lead to the same model.

The question is therefore: can we derive a degenerate model from fermion calculations with level dependent degeneracies?

\subsection{Deriving a degenerate model}

For $i\in I$ and $n=1,\dots,d_i$, we denote by $c_i^{n\dag}$ and $c_i^n$ the creation and annihilation operators associated to the $n$th degenerate sub-level of level $i$.
To ensure a coherent degenerate model we have to assume the following commutation rules
\begin{equation*}
\{c_i^n,c_j^m\}=\{c_i^{n\dag},c_j^{m\dag}\}=0, \hspace{1cm} \{c_i^n,c_j^{m\dag}\} = \delta_{i,j}\delta_{n,m}.
\end{equation*}
The energy $\epsilon_i$ does not depend on the sub-level index $n$. 
Since the levels are exactly the same (and not just have the same energy), the associated wave functions are the same and therefore the dipolar moment entries
only depend on the levels and not on the sub-level. 
We still denote them by $\bfM_{ij}$.
The Bloch equations governing $\rho_{ij}^{nm}=\langle c_j^{m\dag}c_i^n \rangle$
is the Full Degenerate Bloch (FDB) model
\begin{equation}
\label{Eq_degenerate}
\ci\hbar\partial_t \rho_{ij}^{nm} = (\epsilon_i-\epsilon_j) \rho_{ij}^{nm} 
+ \Re\bfE(t) \cdot \sum_{k\in I} \sum_{p=1}^{d_k} (\bfM_{ik} \rho_{kj}^{pm} 
- \bfM_{kj} \rho_{ik}^{np}).
\end{equation}
We notice that natural quantities that arise in the right handside are
\begin{equation*}
\rho_{kj}^{+m} = \sum_{p=1}^{d_k} \rho_{kj}^{pm}, 
\hspace{1cm}
\rho_{ik}^{n+}= \sum_{p=1}^{d_k} \rho_{ik}^{np},
\end{equation*}
and we write the equation for $\rho_{ij}^{n+}$
\begin{eqnarray*}
\ci\hbar\partial_t \rho_{ij}^{n+} 
& = & \ci\hbar\partial_t \sum_{m=1}^{d_j} \rho_{ij}^{nm} \\
& = & (\epsilon_i-\epsilon_j) \sum_{m=1}^{d_j} \rho_{ij}^{nm} 
+ \Re\bfE(t) \cdot \sum_{k\in I} (\bfM_{ik} \sum_{m=1}^{d_j} \rho_{kj}^{+m} 
- \bfM_{kj} \sum_{m=1}^{d_j} \rho_{ik}^{n+}) \\
& = & (\epsilon_i-\epsilon_j) \rho_{ij}^{n+}
+ \Re\bfE(t) \cdot \sum_{k\in I} (\bfM_{ik} \rho_{kj}^{++} - \bfM_{kj} d_j \rho_{ik}^{n+}),
\end{eqnarray*}
where we have introduced
\begin{equation*}
\rho_{ij}^{++} = \sum_{n=1}^{d_i} \sum_{m=1}^{d_j} \rho_{ij}^{nm}
\end{equation*}
and we obtain a closed set of equations for the density matrix $\rho^{++}$, which is the Condensed Degenerate Bloch (CDB) model
\begin{eqnarray*}
\ci\hbar\partial_t \rho_{ij}^{++} 
& = & \ci\hbar\partial_t \sum_{n=1}^{d_i} \rho_{ij}^{n+} \\
& = & (\epsilon_i-\epsilon_j) \sum_{n=1}^{d_i} \rho_{ij}^{n+} 
+ \Re\bfE(t) \cdot
\sum_{k\in I} (\bfM_{ik} \sum_{n=1}^{d_i} \rho_{kj}^{++} 
- \bfM_{kj} d_j \sum_{n=1}^{d_i} \rho_{ik}^{n+}) \\
& = & (\epsilon_i-\epsilon_j) \rho_{ij}^{++}
+ \Re\bfE(t) \cdot \sum_{k\in I} 
(\bfM_{ik} d_i \rho_{kj}^{++} - \bfM_{kj} d_j \rho_{ik}^{++}).
\end{eqnarray*}
It is not exactly the same equation as for non degenerate levels since the degeneracies occur in the equation coefficients.
If we set
\begin{equation*}
\sigma_{ij} = \frac{\rho_{ij}^{++}}{\sqrt{d_id_j}},
\hspace{1cm}
\bfN_{ij} = \bfM_{ij}\sqrt{d_id_j},
\end{equation*} 
we recover the usual Bloch equation
\begin{equation}
\label{Eq_sigma}
\ci\hbar\partial_t \sigma_{ij} = (\epsilon_i-\epsilon_j) \sigma_{ij} 
+ \Re\bfE(t) \cdot \sum_{k\in I} (\bfN_{ik} \sigma_{kj} - \bfN_{kj} \sigma_{ik}).
\end{equation}
This ensures in particular that $\sigma$ (and therefore $\rho^{++}$) defines a non negative operator if this is valid at the initial time.

\subsection{Boundedness of degenerate levels}

The diagonal elements are $\sigma_{ii}=\rho_{ii}^{++}/d_i$ and we would expect this model to ensure $\sigma_{ii}(t)\leq1$ (i.e. $\rho_{ii}^{++}\leq d_i$) for all time.
By the same arguments as above applied on equation \eqref{Eq_sigma}, this is true if this hold at the initial time. 
The problem is that this condition is not natural when dealing with the variables of the FDB model \eqref{Eq_degenerate}.
Indeed the diagonal entries of the CDB model are not the sum of the populations of a given level but also include intra-level coherences. 
An other consequence of the non negativeness of $\rho$ is that $|\rho_{ij}|\leq\sqrt{\rho_{ii}\rho_{jj}}$ (for the variables of equation \eqref{Eq_Bloch1}, see \cite{Bidegaray-Bourgeade-Reignier01}).
If plugged in the definition of $\sigma_{ii}$ this yields
\begin{eqnarray*}
\sigma_{ii} 
& \leq & \frac1{d_i} \sum_{n=1}^{d_i} \sum_{m=1}^{d_i} \rho_{ii}^{nm} 
\leq \frac1{d_i} \sum_{n=1}^{d_i} \sum_{m=1}^{d_i} \sqrt{\rho_{ii}^{nn}}
\sqrt{\rho_{ii}^{mm}} \\
& \leq & \frac1{d_i} \left(\sum_{n=1}^{d_i} \sqrt{\rho_{ii}^{nn}}\right)
\left(\sum_{m=1}^{d_i} \sqrt{\rho_{ii}^{mm}}\right) \leq d_i.
\end{eqnarray*}
Since there can be configurations where $\sigma_{ii}=d_i$, the only way to ensure that $\sigma_{ii}(0)\leq1$ is to impose that intra-level coherences are initially zero.
This certainly is valid in most experimental situations.
Under this vanishing condition, the CDB model preserves the natural property
$\rho_{ii}^{++}\leq d_i$.
In the opposite situation, it is not possible to preserve the property
$\rho_{ii}^{++}\leq d_i^2$ which reverts to preserve $\sigma_{ii}\leq d_i$ and is the same problem as deriving a degenerate model based on a boson derivation of Bloch equations, the initial problem that we have eluded.

\section{The case of two species of electrons}
\label{Sec_2fermion}

\subsection{Conduction--valence notations}

To be closer to the above computations, we will first present the model of two species of fermions considering electron operators (and not electron and hole operators, as in \cite{Gehrig-Hess02}). For $i\in \Ic$ (a set of integers, indexing conduction electrons), we keep the notations $c_i^\dag$ and $c_i$ for the creation and annihilation operators associated to conduction electrons.
For $i\in \Iv$ (indexing valence electrons), we will use the non conventional notation $v_i^\dag$ and $v_i$ for the creation and annihilation operators associated to valence electrons.
Along with the already defined commutation rules for conduction electrons
\begin{equation*}
\{c_i,c_j\}=\{c_i^\dag,c_j^\dag\}=0, \hspace{1cm} \{c_i,c_j^\dag\} = \delta_{i,j},
\end{equation*}
we have the same rules for valence electrons
\begin{equation*}
\{v_i,v_j\}=\{v_i^\dag,v_j^\dag\}=0, \hspace{1cm} \{v_i,v_j^\dag\} = \delta_{i,j}
\end{equation*}
and commutation rules between the two species
\begin{equation*}
[c_i,v_j]=[c_i^\dag,v_j^\dag]=[c_i,v_j^\dag]=[c_i^\dag,v_j]=0.
\end{equation*}
We now consider the Hamiltonians
\begin{equation*}
H^{\rm c} = \sum_{k\in \Ic} \epsilon_k^{\rm c} c_k^\dag c_k,
\hspace{1cm}
H^{\rm v} = \sum_{k\in \Iv} \epsilon_k^{\rm v} v_k^\dag v_k,
\end{equation*}
\begin{eqnarray*}
H^{\rm Lc} 
& = & \frac12 \sum_{(k,l)\in (\Ic)^2} ( \bfE(t) \cdot \bfM_{kl}^{\rm c} c_k^\dag c_l + \bfE^*(t) \cdot \bfM_{kl}^{\rm c*} c_l^\dag c_k), \\
H^{\rm Lv} 
& = & \frac12 \sum_{(k,l)\in (\Iv)^2} ( \bfE(t) \cdot \bfM_{kl}^{\rm v} v_k^\dag v_l + \bfE^*(t) \cdot \bfM_{kl}^{\rm v*} v_l^\dag v_k), \\
H^{\rm Lcv} 
& = & \sum_{(k,l)\in \Ic\times \Iv} ( \bfE(t) \cdot \bfM_{kl}^{\rm cv} c_k^\dag v_l + \bfE^*(t) \cdot \bfM_{kl}^{\rm cv*} v_l^\dag c_k).
\end{eqnarray*}
The density matrices we are interested in are intra-band densities
\begin{equation*}
\rho_{ij}^{\rm c} = \langle c_j^\dag c_i \rangle,
\hspace{1cm}
\rho_{ij}^{\rm v} = \langle v_j^\dag v_i \rangle,
\end{equation*}
and inter-band densities
\begin{equation*}
\rho_{ij}^{\rm cv} = \langle v_j^\dag c_i \rangle,
\hspace{1cm}
\rho_{ij}^{\rm vc} = \langle c_j^\dag v_i \rangle.
\end{equation*}
We will not write the equations for $\rho^{\rm vc}$ but an appropriate definition of $\bfM^{\rm vc}$ ensures that
\begin{equation*}
\rho^{\rm tot}=\left(\begin{array}{cc}
\rho^{\rm c} & \rho^{\rm cv} \\
\rho^{\rm cv*} & \rho^{\rm v} 
\end{array}\right)
\end{equation*}
is Hermitian.

We now have to compute the commutators of these matrices with the three above Hamiltonians. This is done in Appendix \ref{App_2fermion}.

\subsection{Towards a two-species Bloch equation}

Gathering all the computations of Appendix \ref{App_2fermion} we obtain
\begin{equation*}
\begin{aligned}
\ci\hbar\partial_t \rho_{ij}^{\rm c} = 
(\epsilon_i^{\rm c} - \epsilon_j^{\rm c}) \rho_{ij}^{\rm c}
& + \Re\bfE(t) \cdot \sum_{k\in \Ic} (\bfM_{ik}^{\rm c} \rho_{kj}^{\rm c}
- \bfM_{kj}^{\rm c} \rho_{ik}^{\rm c}) \\
&+ \bfE(t) \cdot \sum_{k\in \Iv} \bfM_{ik}^{\rm cv} \rho_{kj}^{\rm vc}
- \bfE^*(t) \cdot \sum_{k\in \Iv} \bfM_{jk}^{\rm cv*} \rho_{ik}^{\rm cv},
\allowdisplaybreaks \\
\ci\hbar\partial_t \rho_{ij}^{\rm v} =
(\epsilon_i^{\rm v} - \epsilon_j^{\rm v}) \rho_{ij}^{\rm v}
&+ \Re\bfE(t) \cdot \sum_{k\in \Iv} (\bfM_{ik}^{\rm v} \rho_{kj}^{\rm v}
- \bfM_{kj}^{\rm v} \rho_{ik}^{\rm v}) \\
&- \bfE(t) \cdot \sum_{k\in \Ic} \bfM_{kj}^{\rm cv} \rho_{ik}^{\rm vc}
- \bfE^*(t) \cdot \sum_{k\in \Ic} \bfM_{ki}^{\rm cv*} \rho_{kj}^{\rm cv},
\allowdisplaybreaks \\
\ci\hbar\partial_t \rho_{ij}^{\rm cv} =
(\epsilon_i^{\rm c}-\epsilon_j^{\rm v}) \rho_{ij}^{\rm cv}
&+ \Re\bfE(t) \cdot (\sum_{k\in \Ic} \bfM_{ik}^{\rm c} \rho_{kj}^{\rm cv} 
- \sum_{k\in \Iv} \bfM_{kj}^{\rm v}\rho_{ik}^{\rm cv}) \\
&+ \bfE(t) \cdot (\sum_{k\in \Iv} \bfM_{ik}^{\rm cv} \rho_{kj}^{\rm v}
- \sum_{k\in \Ic} \bfM_{kj}^{\rm cv} \rho_{ik}^{\rm c}).
\end{aligned}
\end{equation*}
If we denote $E^{\rm c}=\diag(\{\epsilon_i^{\rm c}\}_{i\in \Ic})$, 
$E^{\rm v}=\diag(\{\epsilon_i^{\rm v}\}_{i\in \Iv})$ and
\begin{equation*}
V(t) = \left(\begin{array}{cc}
E^{\rm c} + \Re\bfE(t)\cdot\bfM^{\rm c} & \bfE(t)\cdot\bfM^{\rm cv} \\
\bfE^*(t)\cdot\bfM^{\rm cv*} & E^{\rm v} + \Re\bfE(t)\cdot\bfM^{\rm v} \\
\end{array}\right),
\end{equation*}
we have the Bloch equation in Liouville form
\begin{equation*}
\ci\hbar\partial_t \rho^{\rm tot} = [V(t),\rho^{\rm tot}].
\end{equation*}
We can in a straightforward way apply our discussion about the degeneracy of levels to this two-species context.

\subsection{Electron--hole formulation}

Since all valence electrons do not play a role in the interaction with conduction
electrons, only those who interact are described and not by their presence but their absence: a hole in the valence band.
We formulate anew the former notations and results, denoting for 
$i\in I^{\rm h}=\Iv$ by $d_i^\dag=v_i$ the creation operator of a hole (annihilation of the corresponding electron) and  $d_i=v_i^\dag$ its annihilation operator. 
The hole energy is the opposite of energy of the corresponding electron 
$\epsilon_i^{\rm h}=-\epsilon_i^{\rm v}$. 
Moreover $d_i^\dag d_i = 1-d_i d_i^\dag = 1 -v_i^\dag v_i$, hence the Hamiltonian
for holes is derived from the Hamiltonian for valence electrons \textit{via}
\begin{equation*}
H^{\rm v} = \sum_{k\in \Iv} \epsilon_k^{\rm v} v_k^\dag v_k
= \sum_{k\in I^{\rm h}} -\epsilon_k^{\rm h} (1-d_k^\dag d_k)
= \sum_{k\in I^{\rm h}} -\epsilon_k^{\rm h} 
+ \sum_{k\in I^{\rm h}} \epsilon_k^{\rm h} d_k^\dag d_k.
\end{equation*}
The first term is a constant that will not play any r\^ole in the commutators, hence we define
\begin{equation*}
H^{\rm h} = \sum_{k\in I^{\rm h}} \epsilon_k^{\rm h} d_k^\dag d_k.
\end{equation*}
The intra-band and inter-band density matrices involving holes are defined by
$\rho_{ij}^{\rm h}=\langle d^\dag_j d_i \rangle = \delta_{ij} - \rho_{ji}^{\rm v}$
and $\rho_{ij}^{\rm ch} = \langle d_j c_i \rangle = \rho_{ij}^{\rm cv}$.
The system for these variables read
\begin{equation}
\label{Eq_Bloch2eh}
\begin{aligned}
\ci\hbar\partial_t \rho_{ij}^{\rm c} = 
(\epsilon_i^{\rm c} - \epsilon_j^{\rm c}) \rho_{ij}^{\rm c}
& + \Re\bfE(t) \cdot \sum_{k\in \Ic} (\bfM_{ik}^{\rm c} \rho_{kj}^{\rm c}
- \bfM_{kj}^{\rm c} \rho_{ik}^{\rm c}) \\
& + \bfE(t) \cdot \sum_{k\in I^{\rm h}} \bfM_{ik}^{\rm ch} \rho_{kj}^{\rm hc}
- \bfE^*(t) \cdot \sum_{k\in I^{\rm h}} \bfM_{jk}^{\rm ch*} \rho_{ik}^{\rm ch},
\allowdisplaybreaks \\
\ci\hbar\partial_t \rho_{ij}^{\rm h} = 
(\epsilon_i^{\rm h} - \epsilon_j^{\rm h}) \rho_{ij}^{\rm h}
& + \Re\bfE(t) \cdot \sum_{k\in I^{\rm h}} (\bfM_{jk}^{\rm v} \rho_{ik}^{\rm h}
- \bfM_{ki}^{\rm v} \rho_{kj}^{\rm h}) \\
& + \bfE(t) \cdot \sum_{k\in \Ic} \bfM_{ki}^{\rm ch} \rho_{jk}^{\rm hc}
- \bfE^*(t) \cdot \sum_{k\in \Ic} \bfM_{kj}^{\rm ch*} \rho_{ki}^{\rm ch},
\allowdisplaybreaks \\
\ci\hbar\partial_t \rho_{ij}^{\rm ch} = 
(\epsilon_i^{\rm c}+\epsilon_j^{\rm h}) \rho_{ij}^{\rm ch}
& + \Re\bfE(t) \cdot (\sum_{k\in \Ic} \bfM_{ik}^{\rm c} \rho_{kj}^{\rm ch} 
- \sum_{k\in I^{\rm h}} \bfM_{kj}^{\rm h}\rho_{ik}^{\rm ch}) \\
& + \bfE(t) \cdot (\sum_{k\in I^{\rm h}} \bfM_{ik}^{\rm ch} (\delta_{jk}-\rho_{jk}^{\rm h})
- \sum_{k\in \Ic} \bfM_{kj}^{\rm ch} \rho_{ik}^{\rm c}).
\end{aligned}
\end{equation}
which has lost the Liouville structure. 
Here we have chosen to denote $\bfM_{ij}^{\rm h}=\bfM_{ij}^{\rm v}$ and 
$\bfM_{ij}^{\rm ch}=\bfM_{ij}^{\rm cv}$ but no other choice would help to recover this structure.
Mathematical results have to be obtained from the electron--electron formulation.
The electron--hole formulation should be only kept for simulation or intuition in introducing new terms such as electron--hole recombination.

\subsection{Vanishing intra-band coherences}

In comparison with the model proposed in \cite{Gehrig-Hess02}, a natural question is now whether we recover their model by imposing vanishing intra-band coherences, i.e. $\rho_{ij}^{\rm c}=\rho_{ij}^{\rm v}=0$ for $i\neq j$, .
In this reference the variables are $n_i^{\rm e} = \rho_{ii}^{\rm c}$,
$n_j^{\rm h} = \rho_{jj}^{\rm h}$ and $p_{ji} = \rho_{ij}^{\rm ch}$ (with a subtlety about coupling with forward and backward propagating optical fields, which we do not separate here).
With this set of variables and vanishing intra-band coherences, the system
\eqref{Eq_Bloch2eh} now reads
\begin{eqnarray*}
\ci\hbar\partial_t n_i^{\rm e} 
& = & \bfE(t) \cdot \sum_{k\in I^{\rm h}} \bfM_{ik}^{\rm ch} p_{ki}^* 
- \bfE^*(t) \cdot \sum_{k\in I^{\rm h}} \bfM_{jk}^{\rm ch*} p_{ki},
\allowdisplaybreaks \\
\ci\hbar\partial_t n_j^{\rm h}  & = & 
\bfE(t) \cdot \sum_{k\in \Ic} \bfM_{kj}^{\rm ch} p_{jk}^* 
- \bfE^*(t) \cdot \sum_{k\in \Ic} \bfM_{kj}^{\rm ch*} p_{jk},
\allowdisplaybreaks \\
\ci\hbar\partial_t p_{ji} & = & 
\begin{aligned}[t]
(\epsilon_i^{\rm c}+\epsilon_j^{\rm h}) p_{ji} 
& + \Re\bfE(t) \cdot (\sum_{k\in \Ic} \bfM_{ik}^{\rm c} p_{jk} 
- \sum_{k\in I^{\rm h}} \bfM_{kj}^{\rm h} p_{ki}) \\
&+ \bfE(t) \cdot \bfM_{ij}^{\rm ch} (1-n_j^{\rm h} - n_i^{\rm e}).
\end{aligned}
\end{eqnarray*}
The difference with the equations announced in \cite{Gehrig-Hess02} (apart from extra terms describing phenomena out of the scope of the present paper) are the terms involving $\bfM^{\rm c}$ and $\bfM^{\rm h}$.

Instead of assuming that intra-band coherences are zero, we could assume that 
$\bfM^{\rm c}$ and $\bfM^{\rm h}$ are zero (or, since this is certainly not true, that
the Hamiltonians $H^{\rm Lc}$ and $H^{\rm Lh}$ should not be taken into account in the derivation), we also then do not recover the equations in \cite{Gehrig-Hess02} because of extra terms in the equation for $p_{ji}$.
Under this only assumption intra-band coherences are not zero, even if taken to be zero at the initial time.

The equations in \cite{Gehrig-Hess02} have therefore lost the Liouville structure and are not suitable for the mathematical analysis. 
The model we propose is far better in this respect.
If assuming zero intra-band coherences is not coherent with  the Liouville structure, we can easily have them very small by simply assuming them to be initially zero and adding in their equations strong damping terms.
As shown in \cite{Bidegaray-Bourgeade-Reignier01}, positiveness results can still be obtained in this context.
Besides the frequencies of $\bfE$ are chosen to be close to the gap frequency, in order to match valence--conduction transition frequencies.
The intra-band coherences are hence not directly excited by such a wave.

\section{Conclusion}

We have derived raw Bloch equations for quantum boxes, i.e. two species of electrons. 
These equations have the Liouville form which allows to prove positiveness results. 
We have also discussed the problem of level degeneracy and derived a condensed degenerate Bloch model in which diagonal entries remain bounded by the corresponding level degeneracy through time evolution, even when the degeneracy order is level-dependent.

Positiveness and boundedness results are necessary to prove similar properties when extra terms are added in these raw equations.
They are also required to prove sharp existence results when coupled with a model for wave propagation (e.g. Maxwell equations) as already done for one species of electrons \cite{Dumas05}.

Future work will consist in introducing with the same care extra contributions to the Hamiltonian to model Coulomb effect, electron--phonon interactions, electron--hole recombination, \dots\ Our goal is to obtain a model close to that of \cite{Gehrig-Hess02} but with anew possible other contributions, and full equations for intra-band coherences.

\appendix

\section{Detail of one species fermion computations}
\label{App_1fermion}

Let us first compute the commutator with the free electron Hamiltonian.
We compute separately the cases $i\neq j$ and $i=j$ but notice afterwards that results reads the same in both cases namely 
\begin{eqnarray*}
\langle[c_i^\dag c_i,H^{\rm e}]\rangle
& = & \sum_k \epsilon_k \langle c_i^\dag c_i c_k^\dag c_k - c_k^\dag c_k c_i^\dag c_i \rangle \\
& = & \epsilon_i \langle c_i^\dag c_i c_i^\dag c_i - c_i^\dag c_i c_i^\dag c_i \rangle = 0, \\
\langle[c_j^\dag c_i,H^{\rm e}]\rangle
& = & \epsilon_i \langle c_j^\dag c_i c_i^\dag c_i - c_i^\dag c_i c_j^\dag c_i \rangle 
+ \epsilon_j \langle c_j^\dag c_i c_j^\dag c_j - c_j^\dag c_j c_j^\dag c_i \rangle
\\
& = & \epsilon_i \langle c_j^\dag c_i \rangle 
- \epsilon_j \langle c_j^\dag c_i \rangle \\
& = & (\epsilon_i - \epsilon_j) \rho_{ij}.
\end{eqnarray*}
Now we compute the commutators with $H^{\rm L}$.
\begin{eqnarray*}
2\langle[c_i^\dag c_i,H^{\rm L}]\rangle
& = & \bfE(t) \cdot \sum_{l\neq i} \bfM_{il} \langle c_i^\dag c_i c_i^\dag c_l 
- c_i^\dag c_l c_i^\dag c_i \rangle \\
&& + \bfE^*(t) \cdot \sum_{l\neq i} \bfM_{il}^* \langle c_i^\dag c_i c_l^\dag c_i 
- c_l^\dag c_i c_i^\dag c_i \rangle \\ 
&& + \bfE(t) \cdot \sum_{k\neq i} \bfM_{ki} \langle c_i^\dag c_i c_k^\dag c_i -
c_k^\dag c_i c_i^\dag c_i \rangle \\
&& + \bfE^*(t) \cdot \sum_{k\neq i} \bfM_{ki}^* \langle c_i^\dag c_i c_i^\dag c_k 
- c_i^\dag c_k c_i^\dag c_i \rangle 
\allowdisplaybreaks \\
& = & \bfE(t) \cdot \sum_{l\neq i} \bfM_{il} \langle c_i^\dag c_l \rangle 
- \bfE^*(t) \cdot \sum_{l\neq i} \bfM_{il}^* \langle c_l^\dag c_i \rangle \\ 
&& - \bfE(t) \cdot \sum_{k\neq i} \bfM_{ki} \langle c_k^\dag c_i \rangle 
+ \bfE^*(t) \cdot \sum_{k\neq i} \bfM_{ki}^* \langle c_i^\dag c_k \rangle 
\allowdisplaybreaks \\
\langle[c_i^\dag c_i,H^{\rm L}]\rangle & = &  \Re\bfE(t) \cdot \sum_{k} \bfM_{ik} \rho_{ki}
- \Re\bfE(t) \cdot \sum_{k} \bfM_{ik}^* \rho_{ik} \allowdisplaybreaks \\
2\langle[c_j^\dag c_i,H^{\rm L}]\rangle
& = & \bfE(t) \cdot \sum_{l\neq j} \bfM_{il} \langle c_j^\dag c_i c_i^\dag c_l 
- c_i^\dag c_l c_j^\dag c_i \rangle \\
&& + \bfE^*(t) \cdot \sum_{l\neq i} \bfM_{jl}^* \langle c_j^\dag c_i c_l^\dag c_j 
- c_l^\dag c_j c_j^\dag c_i \rangle \\
&& + \bfE(t) \cdot \sum_{k\neq i} \bfM_{kj} \langle c_j^\dag c_i c_k^\dag c_j 
- c_k^\dag c_j c_j^\dag c_i \rangle \\
&& + \bfE^*(t) \cdot \sum_{k\neq j} \bfM_{ki}^* \langle c_j^\dag c_i c_i^\dag c_k 
- c_i^\dag c_k c_j^\dag c_i \rangle \\
&& + \bfE(t) \cdot \bfM_{ij} \langle c_j^\dag c_i c_i^\dag c_j 
- c_i^\dag c_j c_j^\dag c_i \rangle \\
&& + \bfE^*(t) \cdot \bfM_{ji}^* \langle c_j^\dag c_i c_i^\dag c_j 
- c_i^\dag c_j c_j^\dag c_i \rangle
\allowdisplaybreaks \\
& = & \bfE(t) \cdot \sum_{l\neq j} \bfM_{il} \langle c_j^\dag (1-c_i^\dag c_i) c_l 
- c_i^\dag c_l c_j^\dag c_i \rangle \\
&& + \bfE^*(t) \cdot \sum_{l\neq i} \bfM_{jl}^* \langle c_j^\dag c_i c_l^\dag c_j 
- c_l^\dag (1-c_j^\dag c_j) c_i \rangle \\
&& + \bfE(t) \cdot \sum_{k\neq i} \bfM_{kj} \langle c_j^\dag c_i c_k^\dag c_j 
- c_k^\dag (1-c_j^\dag c_j) c_i \rangle \\
&& + \bfE^*(t) \cdot \sum_{k\neq j} \bfM_{ki}^* \langle c_j^\dag (1-c_i^\dag c_i) c_k 
- c_i^\dag c_k c_j^\dag c_i \rangle \\
&& + \bfE(t) \cdot \bfM_{ij} \langle c_j^\dag (1-c_i^\dag c_i) c_j 
- c_i^\dag (1-c_j^\dag c_j) c_i \rangle \\
&& + \bfE^*(t) \cdot \bfM_{ji}^* \langle c_j^\dag (1-c_i^\dag c_i) c_j 
- c_i^\dag (1-c_j^\dag c_j) c_i \rangle
\allowdisplaybreaks \\
& = & \bfE(t) \cdot \sum_{l\neq j} \bfM_{il} \langle c_j^\dag c_l \rangle 
- \bfE^*(t) \cdot \sum_{l\neq i} \bfM_{jl}^* \langle c_l^\dag c_i \rangle \\
&& - \bfE(t) \cdot \sum_{k\neq i} \bfM_{kj} \langle c_k^\dag c_i \rangle 
+ \bfE^*(t) \cdot \sum_{k\neq j} \bfM_{ki}^* \langle c_j^\dag c_k \rangle \\
&& + \bfE(t) \cdot \bfM_{ij} \langle c_j^\dag c_j - c_i^\dag c_i \rangle 
+ \bfE^*(t) \cdot \bfM_{ji}^* \langle c_j^\dag  c_j - c_i^\dag c_i \rangle
\allowdisplaybreaks \\
\langle[c_j^\dag c_i,H^{\rm L}]\rangle
& = & \Re\bfE(t) \cdot \sum_{k} \bfM_{ik} \rho_{kj}
- \Re\bfE(t) \cdot \sum_{k} \bfM_{kj} \rho_{ik}.
\end{eqnarray*}

\section{Boson computations}
\label{App_boson}

If electrons are considered as bosons, or in other words without Pauli exception rule, the definitions for the operators, Hamiltonian and density matrix are the same as for fermions, but the computations differ because of commutation rules between creation and annihilation operators which are now
\begin{equation*}
[c_i,c_j]=[c_i^\dag,c_j^\dag]=0, \hspace{1cm} [c_i,c_j^\dag] = \delta_{i,j}.
\end{equation*}

We still have $\langle[c_i^\dag c_i,H^{\rm e}]\rangle=0$.
We first compute the commutators of the density matrix elements with the electron Hamiltonian
\begin{eqnarray*}
\langle[c_j^\dag c_i,H^{\rm e}]\rangle
& = & \sum_k \epsilon_k \langle c_j^\dag c_i c_k^\dag c_k - c_k^\dag c_k c_j^\dag c_i \rangle \\
& = & \epsilon_i \langle c_j^\dag c_i c_i^\dag c_i - c_i^\dag c_i c_j^\dag c_i \rangle 
+ \epsilon_j \langle c_j^\dag c_i c_j^\dag c_j - c_j^\dag c_j c_j^\dag c_i \rangle
\\
& = & \epsilon_i \langle c_j^\dag (1+c_i^\dag c_i) c_i - c_i^\dag c_i c_j^\dag c_i \rangle 
+ \epsilon_j \langle c_j^\dag c_i c_j^\dag c_j - c_j^\dag (1+c_j^\dag c_j) c_i \rangle
\\
& = & \epsilon_i \langle c_j^\dag c_i \rangle 
- \epsilon_j \langle c_j^\dag c_i \rangle
\\
& = & (\epsilon_i-\epsilon_j) \rho_{ij}.
\end{eqnarray*}
Then we compute the commutators of the density matrix elements with the laser Hamiltonian
\begin{eqnarray*}
2\langle[c_i^\dag c_i,H^{\rm L}]\rangle
& = & \bfE(t) \cdot \sum_{l\neq i} \bfM_{il} \langle c_i^\dag c_i c_i^\dag c_l 
- c_i^\dag c_l c_i^\dag c_i \rangle \\
&& + \bfE^*(t) \cdot \sum_{l\neq i} \bfM_{il}^* \langle c_i^\dag c_i c_l^\dag c_i 
- c_l^\dag c_i c_i^\dag c_i \rangle \\ 
&& + \bfE(t) \cdot \sum_{k\neq i} \bfM_{ki} \langle c_i^\dag c_i c_k^\dag c_i -
c_k^\dag c_i c_i^\dag c_i \rangle \\
&& + \bfE^*(t) \cdot \sum_{k\neq i} \bfM_{ki}^* \langle c_i^\dag c_i c_i^\dag c_k 
- c_i^\dag c_k c_i^\dag c_i \rangle 
\allowdisplaybreaks \\
& = & \bfE(t) \cdot \sum_{l\neq i} \bfM_{il} \langle c_i^\dag (1+c_i^\dag c_i) c_l 
- c_i^\dag c_l c_i^\dag c_i \rangle \\
&& + \bfE^*(t) \cdot \sum_{l\neq i} \bfM_{il}^* \langle c_i^\dag c_i c_l^\dag c_i 
- c_l^\dag (1+c_i^\dag c_i) c_i \rangle \\
&& + \bfE(t) \cdot \sum_{k\neq i} \bfM_{ki} \langle c_i^\dag c_i c_k^\dag c_i 
- c_k^\dag (1+c_i^\dag c_i) c_i \rangle \\
&& + \bfE^*(t) \cdot \sum_{k\neq i} \bfM_{ki}^* \langle c_i^\dag (1+c_i^\dag c_i) c_k 
- c_i^\dag c_k c_i^\dag c_i \rangle
\allowdisplaybreaks \\
& = & \bfE(t) \cdot \sum_{l\neq i} \bfM_{il} \langle c_i^\dag c_l \rangle
- \bfE^*(t) \cdot \sum_{l\neq i} \bfM_{il}^* \langle c_l^\dag c_i \rangle \\
&& - \bfE(t) \cdot \sum_{k\neq i} \bfM_{ki} \langle c_k^\dag c_i \rangle
+ \bfE^*(t) \cdot \sum_{k\neq i} \bfM_{ki}^* \langle c_i^\dag c_k \rangle 
\allowdisplaybreaks \\
& = & \bfE(t) \cdot \sum_{k\neq i} \bfM_{ik} \rho_{ki}
- \bfE^*(t) \cdot \sum_{k\neq i} \bfM_{ki} \rho_{ik} \\
&& - \bfE(t) \cdot \sum_{k\neq i} \bfM_{ki} \rho_{ik}
+ \bfE^*(t) \cdot \sum_{k\neq i} \bfM_{ik} \rho_{ki},
\allowdisplaybreaks \\
\langle[c_i^\dag c_i,H^{\rm L}]\rangle & = & \Re\bfE(t) \cdot \sum_{k} \bfM_{ik} \rho_{ki}
- \Re\bfE(t) \cdot \sum_{k} \bfM_{ik}^* \rho_{ik}, 
\allowdisplaybreaks \\
2\langle[c_j^\dag c_i,H^{\rm L}]\rangle
& = & \bfE(t) \cdot \sum_{l\neq j} \bfM_{il} \langle c_j^\dag c_i c_i^\dag c_l 
- c_i^\dag c_l c_j^\dag c_i \rangle \\
&& + \bfE^*(t) \cdot \sum_{l\neq i} \bfM_{jl}^* \langle c_j^\dag c_i c_l^\dag c_j 
- c_l^\dag c_j c_j^\dag c_i \rangle \\
&& + \bfE(t) \cdot \sum_{k\neq i} \bfM_{kj} \langle c_j^\dag c_i c_k^\dag c_j 
- c_k^\dag c_j c_j^\dag c_i \rangle \\
&& + \bfE^*(t) \cdot \sum_{k\neq j} \bfM_{ki}^* \langle c_j^\dag c_i c_i^\dag c_k 
- c_i^\dag c_k c_j^\dag c_i \rangle \\
&& + \bfE(t) \cdot \bfM_{ij} \langle c_j^\dag c_i c_i^\dag c_j 
- c_i^\dag c_j c_j^\dag c_i \rangle \\
&& + \bfE^*(t) \cdot \bfM_{ji}^* \langle c_j^\dag c_i c_i^\dag c_j 
- c_i^\dag c_j c_j^\dag c_i \rangle
\allowdisplaybreaks \\
& = & \bfE(t) \cdot \sum_{l\neq j} \bfM_{il} \langle c_j^\dag (1+c_i^\dag c_i) c_l 
- c_i^\dag c_l c_j^\dag c_i \rangle \\
&& + \bfE^*(t) \cdot \sum_{l\neq i} \bfM_{jl}^* \langle c_j^\dag c_i c_l^\dag c_j 
- c_l^\dag (1+c_j^\dag c_j) c_i \rangle \\
&& + \bfE(t) \cdot \sum_{k\neq i} \bfM_{kj} \langle c_j^\dag c_i c_k^\dag c_j 
- c_k^\dag (1+c_j^\dag c_j) c_i \rangle \\
&& + \bfE^*(t) \cdot \sum_{k\neq j} \bfM_{ki}^* \langle c_j^\dag (1+c_i^\dag c_i) c_k 
- c_i^\dag c_k c_j^\dag c_i \rangle \\
&& + \bfE(t) \cdot \bfM_{ij} \langle c_j^\dag (1+c_i^\dag c_i) c_j 
- c_i^\dag (1+c_j^\dag c_j) c_i \rangle \\
&& + \bfE^*(t) \cdot \bfM_{ji}^* \langle c_j^\dag (1+c_i^\dag c_i) c_j 
- c_i^\dag (1+c_j^\dag c_j) c_i \rangle
\allowdisplaybreaks \\
& = & \bfE(t) \cdot \sum_{l\neq j} \bfM_{il} \langle c_j^\dag c_l \rangle 
- \bfE^*(t) \cdot \sum_{l\neq i} \bfM_{jl}^* \langle c_l^\dag c_i \rangle \\
&& - \bfE(t) \cdot \sum_{k\neq i} \bfM_{kj} \langle c_k^\dag c_i \rangle 
+ \bfE^*(t) \cdot \sum_{k\neq j} \bfM_{ki}^* \langle c_j^\dag c_k \rangle \\
&& + \bfE(t) \cdot \bfM_{ij} \langle c_j^\dag c_j - c_i^\dag c_i \rangle 
+ \bfE^*(t) \cdot \bfM_{ji}^* \langle c_j^\dag c_j - c_i^\dag c_i \rangle
\allowdisplaybreaks \\
& = & \bfE(t) \cdot \sum_{k\neq j} \bfM_{ik} \rho_{kj}
- \bfE^*(t) \cdot \sum_{k\neq i} \bfM_{kj} \rho_{ik} \\
&& - \bfE(t) \cdot \sum_{k\neq i} \bfM_{kj} \rho_{ik} 
+ \bfE^*(t) \cdot \sum_{k\neq j} \bfM_{ik} \rho_{kj} \\
&& + \bfE(t) \cdot \bfM_{ij} (\rho_{jj} - \rho_{ii})
+ \bfE^*(t) \cdot \bfM_{ij} (\rho_{jj} - \rho_{ii}),
\allowdisplaybreaks \\
\langle[c_j^\dag c_i,H^{\rm L}]\rangle
& = & \Re\bfE(t) \cdot \sum_{k} \bfM_{ik} \rho_{kj}
- \Re\bfE(t) \cdot \sum_{k} \bfM_{kj} \rho_{ik}.
\end{eqnarray*}

Although calculations are different, we find the same results as in the boson case. 
If we consider the Pauli exception rule or not we obtain the same raw Bloch equations. 

\section{Detail of two-species fermion computations}
\label{App_2fermion}

Because conduction and valence operators commute, we clearly have
\begin{equation*}
\langle[c_j^\dag c_i,H^{\rm v}]\rangle 
= \langle[v_j^\dag v_i,H^{\rm c}]\rangle 
= \langle[c_j^\dag c_i,H^{\rm Lv}]\rangle 
= \langle[v_j^\dag v_i,H^{\rm Lc}]\rangle 
= 0.
\end{equation*}
Besides we have already computed in the one-species case
\begin{equation*}
\langle[c_j^\dag c_i,H^{\rm c}]\rangle 
= (\epsilon_i^{\rm c} - \epsilon_j^{\rm c}) \rho_{ij}^{\rm c},
\hspace{1cm}
\langle[v_j^\dag v_i,H^{\rm v}]\rangle 
= (\epsilon_i^{\rm v} - \epsilon_j^{\rm v}) \rho_{ij}^{\rm v},
\end{equation*}
\begin{eqnarray*}
\langle[c_j^\dag c_i,H^{\rm Lc}]\rangle
& = & \Re\bfE(t) \cdot \sum_{k\in \Ic} (\bfM_{ik}^{\rm c} \rho_{kj}^{\rm c}
- \bfM_{kj}^{\rm c} \rho_{ik}^{\rm c}),
\allowdisplaybreaks \\
\langle[v_j^\dag v_i,H^{\rm Lv}]\rangle
& = & \Re\bfE(t) \cdot \sum_{k\in \Iv} (\bfM_{ik}^{\rm v} \rho_{kj}^{\rm v}
- \bfM_{kj}^{\rm v} \rho_{ik}^{\rm v}).
\end{eqnarray*}
We now compute
\begin{eqnarray*}
\langle[v_j^\dag c_i,H^{\rm c}]\rangle
& = & \sum_{k\in \Ic} \epsilon_k^{\rm c} \langle v_j^\dag c_i c_k^\dag c_k 
- c_k^\dag c_k v_j^\dag c_i\rangle
\allowdisplaybreaks \\
& = & \epsilon_i^{\rm c} \langle v_j^\dag c_i c_i^\dag c_i 
- c_i^\dag c_i v_j^\dag c_i\rangle
= \epsilon_i^{\rm c} \langle v_j^\dag c_i \rangle 
= \epsilon_i^{\rm c} \rho_{ij}^{\rm cv},
\allowdisplaybreaks \\
\langle[v_j^\dag c_i,H^{\rm v}]\rangle
& = & \sum_{k\in \Iv} \epsilon_k^{\rm v} \langle v_j^\dag c_i v_k^\dag v_k 
- v_k^\dag v_k v_j^\dag c_i\rangle
\allowdisplaybreaks \\
& = & \epsilon_j^{\rm v} \langle v_j^\dag c_i v_j^\dag v_j 
- v_j^\dag v_j v_j^\dag c_i\rangle 
= - \epsilon_j^{\rm v} \langle v_j^\dag c_i\rangle 
= - \epsilon_j^{\rm v} \rho_{ij}^{\rm cv}.
\end{eqnarray*}
These computations are valid also for $i=j$, which is a case which has to be treated apart for the following computations.
\begin{eqnarray*}
\langle[c_i^\dag c_i,H^{\rm Lcv}]\rangle
& = & \bfE(t) \cdot \sum_{(k,l)\in \Ic\times \Iv} \bfM_{kl}^{\rm cv} 
\langle c_i^\dag c_i c_k^\dag v_l - c_k^\dag v_l c_i^\dag c_i \rangle \\
&& + \bfE^*(t) \cdot \sum_{(k,l)\in \Ic\times \Iv} \bfM_{kl}^{\rm cv*} 
\langle c_i^\dag c_i v_l^\dag c_k - v_l^\dag c_k c_i^\dag c_i \rangle 
\allowdisplaybreaks \\
& = & \bfE(t) \cdot \sum_{l\in \Iv} \bfM_{il}^{\rm cv} 
\langle c_i^\dag c_i c_i^\dag v_l - c_i^\dag v_l c_i^\dag c_i \rangle \\
&& + \bfE^*(t) \cdot \sum_{l\in \Iv} \bfM_{il}^{\rm cv*} 
\langle c_i^\dag c_i v_l^\dag c_i - v_l^\dag c_i c_i^\dag c_i \rangle 
\allowdisplaybreaks \\
& = & \bfE(t) \cdot \sum_{l\in \Iv} \bfM_{il}^{\rm cv} \langle c_i^\dag v_l \rangle 
- \bfE^*(t) \cdot \sum_{l\in \Iv} \bfM_{il}^{\rm cv*} \langle v_l^\dag c_i \rangle
\allowdisplaybreaks \\
& = & \bfE(t) \cdot \sum_{l\in \Iv} \bfM_{il}^{\rm cv} \rho_{li}^{\rm vc}
- \bfE^*(t) \cdot \sum_{l\in \Iv} \bfM_{il}^{\rm cv*} \rho_{il}^{\rm cv}, 
\allowdisplaybreaks \\
\langle[c_j^\dag c_i,H^{\rm Lcv}]\rangle
& = & \bfE(t) \cdot \sum_{(k,l)\in \Ic\times \Iv} \bfM_{kl}^{\rm cv} 
\langle c_j^\dag c_i c_k^\dag v_l - c_k^\dag v_l c_j^\dag c_i \rangle \\
&& + \bfE^*(t) \cdot \sum_{(k,l)\in \Ic\times \Iv} \bfM_{kl}^{\rm cv*} 
\langle c_j^\dag c_i v_l^\dag c_k - v_l^\dag c_k c_j^\dag c_i \rangle
\allowdisplaybreaks \\
& = & \bfE(t) \cdot \sum_{l\in \Iv} \bfM_{il}^{\rm cv} 
\langle c_j^\dag c_i c_i^\dag v_l - c_i^\dag v_l c_j^\dag c_i \rangle \\
&& + \bfE(t) \cdot \sum_{l\in \Iv} \bfM_{jl}^{\rm cv} 
\langle c_j^\dag c_i c_j^\dag v_l - c_j^\dag v_l c_j^\dag c_i \rangle \\
&& + \bfE^*(t) \cdot \sum_{l\in \Iv} \bfM_{il}^{\rm cv*} 
\langle c_j^\dag c_i v_l^\dag c_i - v_l^\dag c_i c_j^\dag c_i \rangle \\
&& + \bfE^*(t) \cdot \sum_{l\in \Iv} \bfM_{jl}^{\rm cv*} 
\langle c_j^\dag c_i v_l^\dag c_j - v_l^\dag c_j c_j^\dag c_i \rangle
\allowdisplaybreaks \\
& = & \bfE(t) \cdot \sum_{l\in \Iv} \bfM_{il}^{\rm cv} \langle c_j^\dag v_l \rangle 
- \bfE^*(t) \cdot \sum_{l\in \Iv} \bfM_{jl}^{\rm cv*} \langle v_l^\dag c_i \rangle
\allowdisplaybreaks \\
& = & \bfE(t) \cdot \sum_{l\in \Iv} \bfM_{il}^{\rm cv} \rho_{lj}^{\rm vc}
- \bfE^*(t) \cdot \sum_{l\in \Iv} \bfM_{jl}^{\rm cv*} \rho_{il}^{\rm cv},
\allowdisplaybreaks \\
\langle[v_i^\dag v_i,H^{\rm Lcv}]\rangle
& = & \bfE(t) \cdot \sum_{(k,l)\in \Ic\times \Iv} \bfM_{kl}^{\rm cv} 
\langle v_i^\dag v_i c_k^\dag v_l - c_k^\dag v_l v_i^\dag v_i \rangle \\
&& + \bfE^*(t) \cdot \sum_{(k,l)\in \Ic\times \Iv} \bfM_{kl}^{\rm cv*} 
\langle v_i^\dag v_i v_l^\dag c_k - v_l^\dag c_k v_i^\dag v_i \rangle \\
\allowdisplaybreaks \\
& = & \bfE(t) \cdot \sum_{k\in \Ic} \bfM_{ki}^{\rm cv} 
\langle v_i^\dag v_i c_k^\dag v_i - c_k^\dag v_i v_i^\dag v_i \rangle \\
&& + \bfE^*(t) \cdot \sum_{k\in \Ic} \bfM_{ki}^{\rm cv*} 
\langle v_i^\dag v_i v_i^\dag c_k - v_i^\dag c_k v_i^\dag v_i \rangle \\
\allowdisplaybreaks \\
& = & - \bfE(t) \cdot \sum_{k\in \Ic} \bfM_{ki}^{\rm cv} \langle c_k^\dag v_i \rangle 
+ \bfE^*(t) \cdot \sum_{k\in \Ic} \bfM_{ki}^{\rm cv*} \langle v_i^\dag c_k \rangle \\
& = & - \bfE(t) \cdot \sum_{k\in \Ic} \bfM_{ki}^{\rm cv} \rho_{ik}^{\rm vc}
+ \bfE^*(t) \cdot \sum_{k\in \Ic} \bfM_{ki}^{\rm cv*} \rho_{ki}^{\rm cv} \\
\allowdisplaybreaks \\
\langle[v_j^\dag v_i,H^{\rm Lcv}]\rangle
& = & \bfE(t) \cdot \sum_{(k,l)\in \Ic\times \Iv} \bfM_{kl}^{\rm cv} 
\langle v_j^\dag v_i c_k^\dag v_l - c_k^\dag v_l v_j^\dag v_i \rangle \\
&& + \bfE^*(t) \cdot \sum_{(k,l)\in \Ic\times \Iv} \bfM_{kl}^{\rm cv*} 
\langle v_j^\dag v_i v_l^\dag c_k - v_l^\dag c_k v_j^\dag v_i \rangle
\allowdisplaybreaks \\
& = & \bfE(t) \cdot \sum_{k\in \Ic} \bfM_{ki}^{\rm cv} 
\langle v_j^\dag v_i c_k^\dag v_i - c_k^\dag v_i v_j^\dag v_i \rangle \\
&& + \bfE(t) \cdot \sum_{k\in \Ic} \bfM_{kj}^{\rm cv} 
\langle v_j^\dag v_i c_k^\dag v_j - c_k^\dag v_j v_j^\dag v_i \rangle \\
&& + \bfE^*(t) \cdot \sum_{k\in \Ic} \bfM_{ki}^{\rm cv*} 
\langle v_j^\dag v_i v_i^\dag c_k - v_i^\dag c_k v_j^\dag v_i \rangle \\
&& + \bfE^*(t) \cdot \sum_{k\in \Ic} \bfM_{kj}^{\rm cv*} 
\langle v_j^\dag v_i v_j^\dag c_k - v_j^\dag c_k v_j^\dag v_i \rangle
\allowdisplaybreaks \\
& = & - \bfE(t) \cdot \sum_{k\in \Ic} \bfM_{kj}^{\rm cv} \langle c_k^\dag v_i \rangle 
+ \bfE^*(t) \cdot \sum_{k\in \Ic} \bfM_{ki}^{\rm cv*} \langle v_j^\dag c_k \rangle
\allowdisplaybreaks \\
& = & - \bfE(t) \cdot \sum_{k\in \Ic} \bfM_{kj}^{\rm cv} \rho_{ik}^{\rm vc}
+ \bfE^*(t) \cdot \sum_{k\in \Ic} \bfM_{ki}^{\rm cv*} \rho_{kj}^{\rm cv}.
\end{eqnarray*}
Once more by different computations we obtain the same result for both cases $i=j$ and $i\neq j$.
Now
\begin{eqnarray*}
2\langle[v_j^\dag c_i,H^{\rm Lc}]\rangle
& = & \bfE(t) \cdot \sum_{(k,l)\in (\Ic)^2} \bfM_{kl}^{\rm c} 
\langle v_j^\dag c_i c_k^\dag c_l - c_k^\dag c_l v_j^\dag c_i \rangle \\
&& + \bfE^*(t) \cdot \sum_{(k,l)\in (\Ic)^2} \bfM_{kl}^{\rm c*} 
\langle v_j^\dag c_i c_l^\dag c_k - c_l^\dag c_k v_j^\dag c_i \rangle 
\allowdisplaybreaks \\
& = & \bfE(t) \cdot \sum_{l\in \Ic} \bfM_{il}^{\rm c} 
\langle v_j^\dag c_i c_i^\dag c_l - c_i^\dag c_l v_j^\dag c_i \rangle \\
&& + \bfE(t) \cdot \sum_{k\in \Ic} \bfM_{ki}^{\rm c} 
\langle v_j^\dag c_i c_k^\dag c_i - c_k^\dag c_i v_j^\dag c_i \rangle \\
&& + \bfE^*(t) \cdot \sum_{l\in \Ic} \bfM_{il}^{\rm c*} 
\langle v_j^\dag c_i c_l^\dag c_i - c_l^\dag c_i v_j^\dag c_i \rangle \\
&& + \bfE^*(t) \cdot \sum_{k\in \Ic} \bfM_{ki}^{\rm c*} 
\langle v_j^\dag c_i c_i^\dag c_k - c_i^\dag c_k v_j^\dag c_i \rangle 
\allowdisplaybreaks \\
& = & 2 \Re\bfE(t) \cdot \sum_{l\in \Ic} \bfM_{il}^{\rm c} \langle v_j^\dag c_l \rangle 
= 2 \Re\bfE(t) \cdot \sum_{k\in \Ic} \bfM_{ik}^{\rm c} \rho_{kj}^{\rm cv}, 
\allowdisplaybreaks \\
2\langle[v_j^\dag c_i,H^{\rm Lv}]\rangle
& = & \bfE(t) \cdot \sum_{(k,l)\in (\Iv)^2} \bfM_{kl}^{\rm v} 
\langle v_j^\dag c_i v_k^\dag v_l - v_k^\dag v_l v_j^\dag c_i \rangle \\
&& + \bfE^*(t) \cdot \sum_{(k,l)\in (\Iv)^2} \bfM_{(k,l)\in (\Iv)^2}^{\rm v*} 
\langle v_j^\dag c_i v_l^\dag v_k - v_l^\dag v_k v_j^\dag c_i \rangle 
\allowdisplaybreaks \\
& = & \bfE(t) \cdot \sum_{l\in \Iv} \bfM_{jl}^{\rm v} 
\langle v_j^\dag c_i v_j^\dag v_l - v_j^\dag v_l v_j^\dag c_i \rangle \\
&& + \bfE(t) \cdot \sum_{k\in \Iv} \bfM_{kj}^{\rm v} 
\langle v_j^\dag c_i v_k^\dag v_j - v_k^\dag v_j v_j^\dag c_i \rangle \\
&& + \bfE^*(t) \cdot \sum_{l\in \Iv} \bfM_{jl}^{\rm v*} 
\langle v_j^\dag c_i v_l^\dag v_j - v_l^\dag v_j v_j^\dag c_i \rangle \\
&& + \bfE^*(t) \cdot \sum_{k\in \Iv} \bfM_{kj}^{\rm v*} 
\langle v_j^\dag c_i v_j^\dag v_k - v_j^\dag v_k v_j^\dag c_i \rangle 
\allowdisplaybreaks \\
& = & - 2 \Re\bfE(t) \cdot \sum_{k\in \Iv} \bfM_{kj}^{\rm v} \langle v_k^\dag c_i \rangle 
= - 2 \Re\bfE(t) \cdot \sum_{k\in \Iv} \bfM_{kj}^{\rm v}\rho_{ik}^{\rm cv}.
\end{eqnarray*}
Last
\begin{eqnarray*}
\langle[v_j^\dag c_i,H^{\rm Lcv}]\rangle
& = & \bfE(t) \cdot \sum_{(k,l)\in \Ic\times \Iv} \bfM_{kl}^{\rm cv} 
\langle v_j^\dag c_i c_k^\dag v_l - c_k^\dag v_l v_j^\dag c_i \rangle \\
&& + \bfE^*(t) \cdot \sum_{(k,l)\in \Ic\times \Iv} \bfM_{kl}^{\rm cv*} 
\langle v_j^\dag c_i v_l^\dag c_k - v_l^\dag c_k v_j^\dag c_i \rangle 
\allowdisplaybreaks \\
& = & \bfE(t) \cdot \sum_{l\in \Iv} \bfM_{il}^{\rm cv} 
\langle v_j^\dag c_i c_i^\dag v_l - c_i^\dag v_l v_j^\dag c_i \rangle \\
&& + \bfE(t) \cdot \sum_{k\in \Ic} \bfM_{kj}^{\rm cv} 
\langle v_j^\dag c_i c_k^\dag v_j - c_k^\dag v_j v_j^\dag c_i \rangle \\
&& - \bfE(t) \cdot \bfM_{ij}^{\rm cv} 
\langle v_j^\dag c_i c_i^\dag v_j - c_i^\dag v_j v_j^\dag c_i \rangle \\
&& + \bfE^*(t) \cdot \sum_{l\in \Iv} \bfM_{il}^{\rm cv*} 
\langle v_j^\dag c_i v_l^\dag c_i - v_l^\dag c_i v_j^\dag c_i \rangle \\
&& + \bfE^*(t) \cdot \sum_{k\in \Ic} \bfM_{kj}^{\rm cv*} 
\langle v_j^\dag c_i v_j^\dag c_k - v_j^\dag c_k v_j^\dag c_i \rangle \\
&& - \bfE^*(t) \cdot \bfM_{ij}^{\rm cv*} 
\langle v_j^\dag c_i v_j^\dag c_i - v_j^\dag c_i v_j^\dag c_i \rangle 
\allowdisplaybreaks \\
& = & \bfE(t) \cdot \sum_{l\in \Iv} \bfM_{il}^{\rm cv} 
\langle v_j^\dag v_l - \delta_{jl} c_i^\dag c_i \rangle \\
&& + \bfE(t) \cdot \sum_{k\in \Ic} \bfM_{kj}^{\rm cv} 
\langle \delta_{ik} v_j^\dag v_j - c_k^\dag c_i \rangle \\
&& - \bfE(t) \cdot \bfM_{ij}^{\rm cv} 
\langle v_j^\dag v_j - c_i^\dag c_i \rangle \\
\allowdisplaybreaks \\
& = & \bfE(t) \cdot \sum_{l\in \Iv} \bfM_{il}^{\rm cv} \langle v_j^\dag v_l \rangle
- \bfE(t) \cdot \sum_{k\in \Ic} \bfM_{kj}^{\rm cv} \langle c_k^\dag c_i \rangle 
\allowdisplaybreaks \\
& = & \bfE(t) \cdot (\sum_{k\in \Iv} \bfM_{ik}^{\rm cv} \rho_{kj}^{\rm v}
- \sum_{k\in \Ic} \bfM_{kj}^{\rm cv} \rho_{ik}^{\rm c}).
\end{eqnarray*}

\end{document}